\documentclass[fleqn,usenatbib]{mnras}
\usepackage{amsmath}
\usepackage{graphicx}
\usepackage{amssymb}
\usepackage{amssymb}
\hypersetup{
 	colorlinks=true,
 	linkcolor=red,
 	citecolor=blue,
 }

\title[Measurement on the cosmic curvature]{Measurement on the cosmic curvature using the Gaussian process method}
\author[Y. Yang et al.]{Yingjie Yang,$^{1}$\thanks{E-mail: yyj@hust.edu.cn} and Yungui Gong,$^{1}$\thanks{Corresponding author. E-mail: yggong@mail.hust.edu.cn} \\
$^1$School of Physics, Huazhong University of Science and Technology, Wuhan 430074, China}

\date{Accepted 2021 April 14. Received 2021 April 14; in original form 2021 February 3.}

\pubyear{2021}

\begin{document}
\maketitle

\begin{abstract}
Inflation predicts that the Universe is spatially flat. The Planck 2018 measurements of the cosmic microwave background anisotropy favour a spatially closed
universe at more than 2$\sigma$ confidence level.
We use model independent methods to study the issue of cosmic curvature.
The method reconstructs the Hubble parameter $H(z)$ from
cosmic chronometers data with the Gaussian process method.
The distance modulus is then calculated with the reconstructed function $H(z)$
and fitted by type Ia supernovae data.
Combining the cosmic chronometers and type Ia supernovae data,
we obtain $\Omega_{k0}h^2=0.102\pm 0.066$ which is consistent with a spatially flat universe at the 2$\sigma$ confidence level.
By adding the redshift space distortions data to the type Ia supernovae data
with a proposed novel model independent method,
we obtain $\Omega_{k0}h^2=0.117^{+0.058}_{-0.045}$
and no deviation from $\Lambda$CDM model is found.
\end{abstract}


\section{Introduction}
The Planck 2018 temperature and polarization measurements
of the cosmic microwave background anisotropy
find that $\Omega_{k0}=-0.044_{-0.015}^{+0.018}$ \citep{Aghanim:2018eyx},
an apparent detection of cosmic curvature over $2\sigma$ confidence level.
A spatially closed universe preferred by the Planck 2018 result
is inconsistent with inflationary prediction of a flat universe.
Unfortunately, the Planck 2018 result assumes a non-flat $\Lambda$CDM model
and it is model dependent which limits the validity of the result.
Due to the strong degeneracy between the cosmic curvature
and dark energy, the value of the cosmic curvature depends
on the dark energy model used in fitting the observational data \citep{Wang:2004jf,Gong:2005de,Gong:2007wx,Clarkson:2007bc,Pan:2010rn,Gong:2010gs,Dossett:2012kd,Witzemann:2017lhi,Ryan:2018aif,Park:2018tgj,Ryan:2019uor}.
Because the spatial curvature of our Universe has profound consequences
for inflation and fundamental physics, it is an outstanding issue in cosmology
and it is necessary to determine the cosmic curvature with model independent method so that the issue of the curvature tension can be better understood.

The Hubble expansion rate $H(z)$ depends on both the background geometry
and dark energy models, so the determination of the cosmic
curvature from the expansion history of the Universe is model dependent.
The model dependence can be alleviated if we can determine
distances and the Hubble expansion rate directly from observations.
Using the Friedmann-Lema\^{\i}tre-Robertson-Walker (FLRW) metric,
a relation between the luminosity distance and the Hubble expansion rate is derived.
Since distances depend on both the Hubble expansion rate and the spatial curvature,
a null test of the cosmic curvature based
on the relation between distances and the Hubble expansion rate was proposed \citep{Clarkson:2007bc,Clarkson:2007pz}.
The null test of the spatial curvature is independent of cosmological model
and gravitational theory. It can determine the spatial curvature,
test the FLRW metric and detect the tension between observational data.
Due to the singularity of the comoving distance $D(z)$ at
the redshift $z=0$, an alternative null test of the spatial curvature which just tests
the flatness of the Universe was proposed \citep{Cai:2015pia}.
With the non-parametric method of reconstructing the Hubble expansion rate from the cosmic chronometers (CCH) data and reconstructing distances from type Ia supernoave (SNe Ia) or baryon acoustic oscillation data,
the null tests of the spatial curvature were applied to determine the cosmic curvature and test the flatness of
the Universe \citep{Cai:2015pia,Shafieloo:2009hi,Li:2014yza,Sapone:2014nna,LHuillier:2016mtc,Yu:2016gmd,Marra:2017pst,Yu:2017iju,Li:2019bbg}.
Instead of reconstructing distances from observational data and
determining the cosmic curvature at each redshift,
we can obtain the constraint on the cosmic curvature by using the $\chi^2$ minimization \citep{Wei:2016xti,Wei:2018cov,Wang:2017lri,Wei:2019uss,Wang:2020dbt}.
However, the value of the Hubble constant $H_0=100h$ km/s/Mpc is needed to reconstruct
$E(z)=H(z)/H_0$, so the result with this methods depends on the value of the Hubble constant.
Interestingly, the reconstructed $E(z)$
can be used to provide model independent direct evidence of cosmic acceleration \citep{Yang:2019fjt}.
Model independent estimate of the cosmic curvature by using weak
lensing galaxy-shear correlations was proposed in \cite{Bernstein:2005en}.
The possibility of using strong lensing systems and quasars
as standard candels to determine the cosmic curvature was also discussed \citep{Rana:2016gha,Qi:2018aio,Wang:2019yob,Liu:2020bzc}.
By combining time delays between strongly lensed images
of time variable sources and the SNe Ia distance,
both the Hubble constant the cosmic curvature were determined
model independently and no deviation from a spatially flat universe is detected \citep{Collett:2019hrr}.
In this paper, we use the Gaussian process (GP) method to
reconstruct the Hubble parameter $H(z)$ from CCH data and fit the combined parameter $\Omega_{k0} h^2$ to SNe Ia data so that the issue of the value of the Hubble constant is avoided.

Combining the measurements of distances and the growth of large structure, we can estimate the cosmic curvature model independently \citep{Mortonson:2009nw}.
The growth rate of matter perturbation can distinguish modified gravity and dark energy models.
In a spatially curved universe, a good approximation of the growth factor is  $f(z)=\Omega_m^\gamma+(\gamma-4/7)\Omega_k$ \citep{Gong:2009sp},
where the growth index $\gamma$ is an indicator of the underlying model.
For the $\Lambda$CDM model, $\gamma\approx 0.545$.
For the Dvali-Gabadadze-Porrati (DGP)
brane-world model \citep{Dvali:2000hr}, $\gamma\approx 0.6875$.
Armed with the analytical expression for the growth factor,
we propose a model independent method to use the observations
of the redshift space distortions (RSD) to constrain the spatial curvature.
With the reconstructed smooth function $H(z)$ from CCH data,
we can reconstruct the growth factor $f(z)$ and the function $f\sigma_8(z)$.
The cosmic curvature and the growth index
are then determined by fitting the reconstructed $f\sigma_8(z)$ to the RSD data.
In the reconstruction process, no cosmological model or
gravitational theory is used, so the measurements of the cosmic
curvature and the growth index from RSD data are model independent.
The combined CCH, SNe Ia and RSD data are used to determine the cosmic curvature
and the growth index.

The paper is organized as follows. In section \ref{sec2},
we apply the model independent method of null tests of the spatial curvature
to test the flatness
by using the CCH and SNe Ia data.
In section \ref{sec3}, we introduce model independent methods of determining the cosmic curvature. A model independent method of
using the observations of RSD to constrain the spatial curvature is proposed.
The model independent method probes not only
the geometry of the Universe but also the underlying theory.
The CCH, SNe Ia and RSD data are then used to constrain the cosmic curvature.
The paper is concluded in section \ref{sec4}.

\section{The null tests of cosmic curvature}
\label{sec2}
From FLRW metric,
\begin{equation}
\label{frweq}
ds^2=-dt^2+a^2(t)\left[\frac{dr^2}{1-Kr^2}+r^2(d\theta^2+\sin^2\theta d\phi^2)\right],
\end{equation}
we get the luminosity distance
\begin{equation}
\label{dleq}
d_L=\frac{1+z}{H_0\sqrt{-\Omega_{k0}}}\sin\left[\sqrt{-\Omega_{k0}}\int_0^z \frac{dz'}{H(z')}\right],
\end{equation}
where $K$ denotes the spatial curvature and $\Omega_{k0}=-K/H_0^2$.
If $\Omega_{k0}>0$, then the function $\sin(x)$ becomes $\sinh(x)$.
From the definition of the luminosity distance \eqref{dleq},
the spatial curvature can be written as \citep{Clarkson:2007bc,Clarkson:2007pz}
\begin{equation}
\label{ok1}
\Omega_{k0}=\frac{[E(z)D'(z)]^2-1}{D(z)^2},
\end{equation}
where the dimensionless comoving distance $D(z)$
is related with the luminosity distance $d_L$ as $D(z)=H_0 d_L/(1+z)$,
$E(z)=H(z)/H_0$ and the prime refers to the derivative
with respect to the redshift $z$.
Once we know the Hubble rate $H(z)$ and the distance $D(z)$ at a redshift $z$,
then from Eq. \eqref{ok1} we can obtain the cosmic curvature $\Omega_{k0}$.
Because Eq. \eqref{ok1} relies on the FLRW metric only and
it is independent of cosmological model and gravitational theory,
so it not only determines the value of $\Omega_{k0}$ model independently,
the constancy of $\Omega_{k0}$ is also a model independent test of FLRW metric \citep{Clarkson:2007pz} and a consistency check for the observational data.
Since  $D(z=0)=0$ brings a singularity at $z=0$,
we can use an alternative model independent null test \citep{Cai:2015pia}
\begin{equation}
\label{ok2}
\mathcal{O}_k(z)=E(z)D'(z)-1.
\end{equation}
A flat universe implies $\mathcal{O}_k(z)=0$, so it can be used
to test the flatness of the universe, but the value of $\Omega_{k0}$ cannot be directly determined from this null test
and it cannot be used to test the FLRW metric and the consistency of observational data.

\begin{figure}
\includegraphics[width=0.9\linewidth]{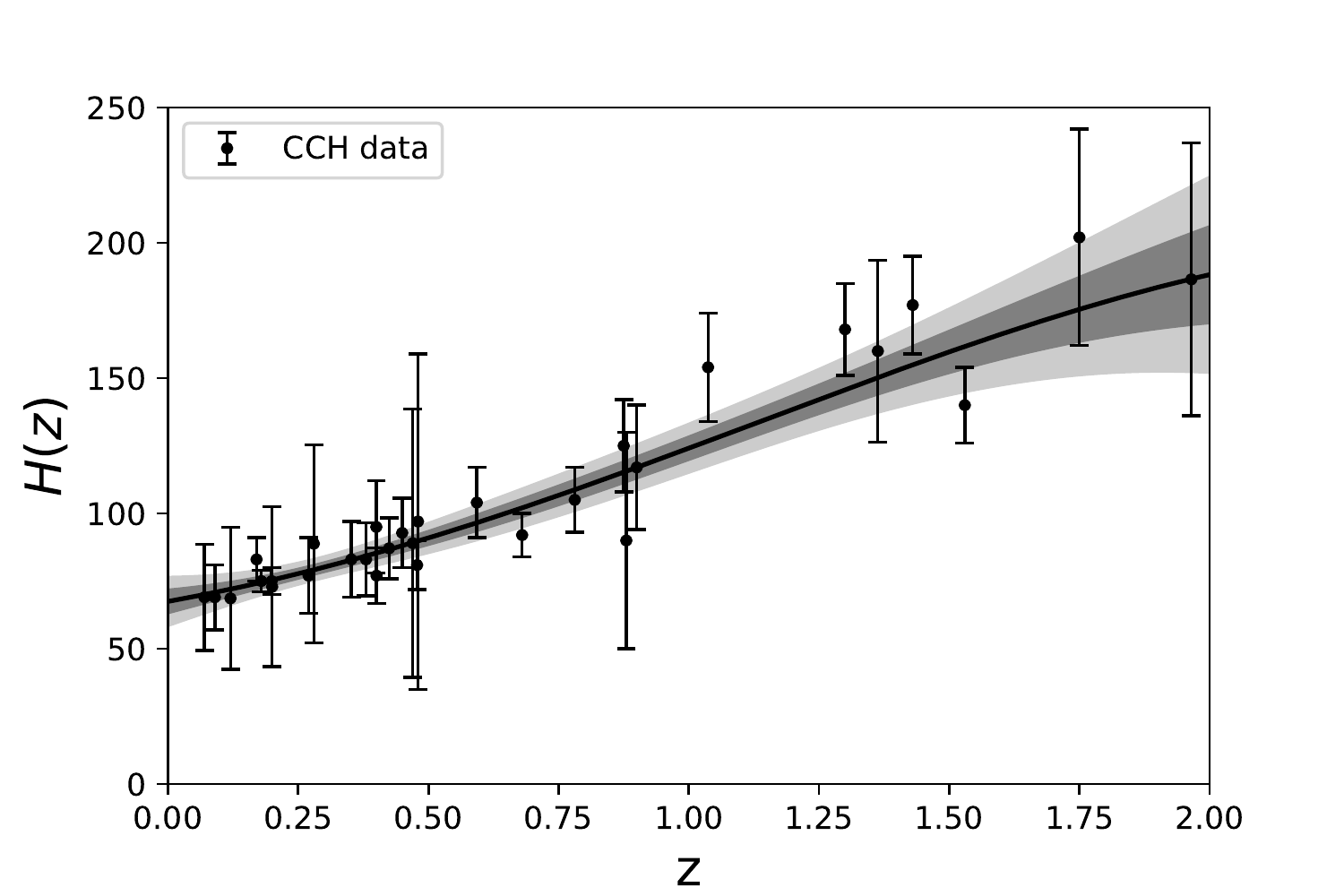}
\caption{The reconstructed Hubble parameter $H(z)$ from the CCH data.
The shaded regions are $1\sigma$ and 2$\sigma$ errors.}
\label{gph}
\end{figure}

To test the flatness of the Universe with the null tests \eqref{ok1} and \eqref{ok2}, we need to know the functions $E(z)$, $D(z)$ and $D'(z)$.
We use the GP method to reconstruct $E(z)$ from CCH data and reconstruct
$D(z)$ and $D'(z)$ from the Pantheon sample of SNe Ia data.
The $H(z)$ data can be obtained with
the CCH method using
the differential redshift time derived from the
spectroscopic differential evolution of passively evolving
galaxies \citep{Jimenez:2001gg}.
In this paper we use the 31 CCH data points \citep{Yang:2019fjt}
which were measured by assuming the BC03 stellar population synthesis model \citep{Simon:2004tf,Stern:2009ep,Zhang:2012mp,Moresco:2012jh,Moresco:2015cya,Moresco:2016mzx,Ratsimbazafy:2017vga}.
These data cover the redshift range up to $z\sim 2$ and assume no particular cosmological model.
We use the public available python package GaPP \citep{Seikel:2012uu} for GP method to reconstruct the Hubble parameter $H(z)$
and the results are shown in Fig. \ref{gph}.
The reconstructed $H(z)$ is in good agreement with previous results \citep{Cai:2015pia,Wei:2016xti,Yu:2017iju,Gomez-Valent:2018hwc,Pinho:2018unz,Haridasu:2018gqm,Yang:2019fjt}.
From this model independent reconstruction,
we get the Hubble constant $H_0=67.46 \pm 4.75$ km/s/Mpc and this value will be used for the null tests below.

\begin{figure}
\includegraphics[width=0.9\linewidth]{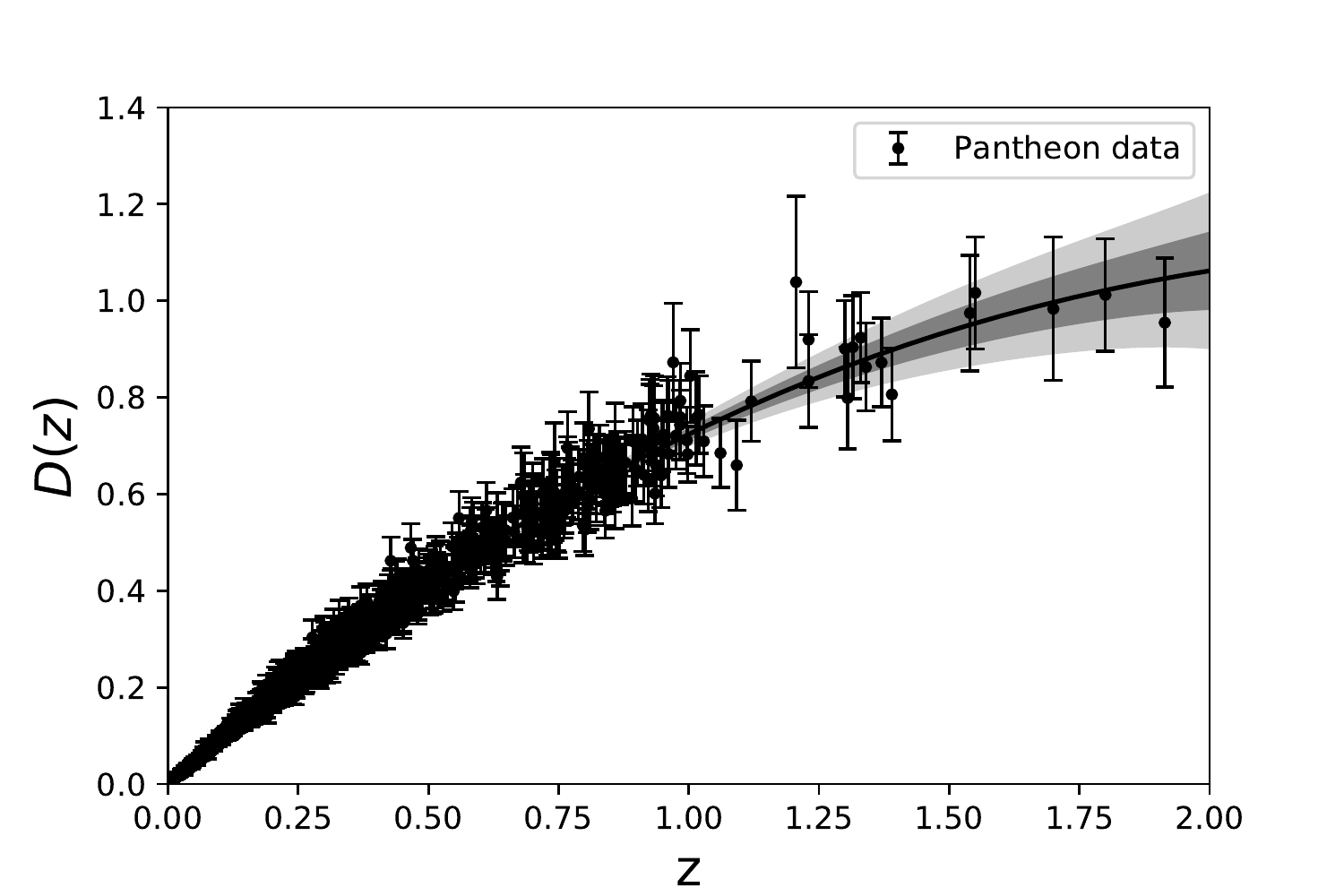}
\caption{The reconstructed dimensionless comoving distance $D(z)$ from the SNe Ia data.
The shaded regions are $1\sigma$ and 2$\sigma$ errors.}
\label{gphd}
\end{figure}

The comoving distance can be measured from SNe Ia observations.
The latest Pantheon sample \citep{Scolnic:2017caz} includes 1048 spectroscopically confirmed SNe Ia with the redshift up to $z\sim 2.3$.
It consists of 279 spectroscopically confirmed SNe Ia with
redshift $0.03<z<0.68$ discovered by the Pan-STARRS1 Medium Deep Survey \citep{Rest:2013mwz},
samples of SNe Ia from the Harvard Smithsonian Center for Astrophysics SN surveys \citep{Hicken:2009df}, the Carnegie SN Project \citep{Stritzinger:2011qd}, the Sloan digital sky survey \citep{Kessler:2009ys} and the SN legacy survey \citep{Conley:2011ku},
and high-$z$ data with the redshift $z > 1.0$ from the Hubble space telescope cluster SN survey \citep{Suzuki:2011hu}, GOODS \citep{Riess:2006fw} and CANDELS/CLASH survey \citep{Rodney:2014twa,Graur:2013msa}. The calibration systematics is reduced substantially by cross-calibrating all of the SN samples.
We reconstruct the comoving distance $D(z)$ from the 1048 Pantheon sample of SNe Ia with the GP method,
and the results are shown in Fig. \ref{gphd}.

\begin{figure}
\includegraphics[width=0.9\linewidth]{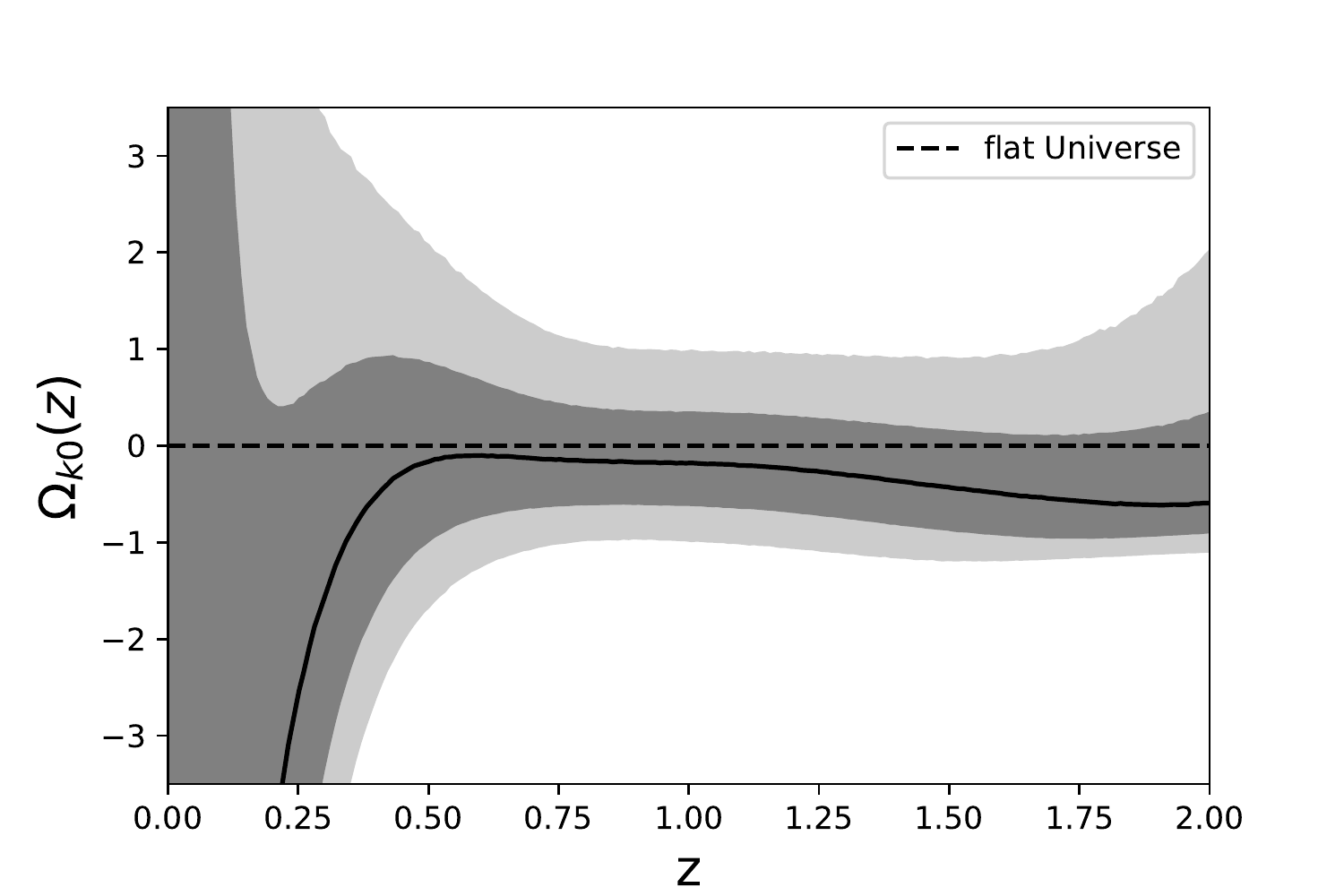}
\caption{The reconstructed $\Omega_{k0}$ from the SNe Ia and CCH data. The shaded areas are the 1$\sigma$ and 2$\sigma$ confidence regions.}
\label{Ok}
\end{figure}

\begin{figure}
\includegraphics[width=0.9\linewidth]{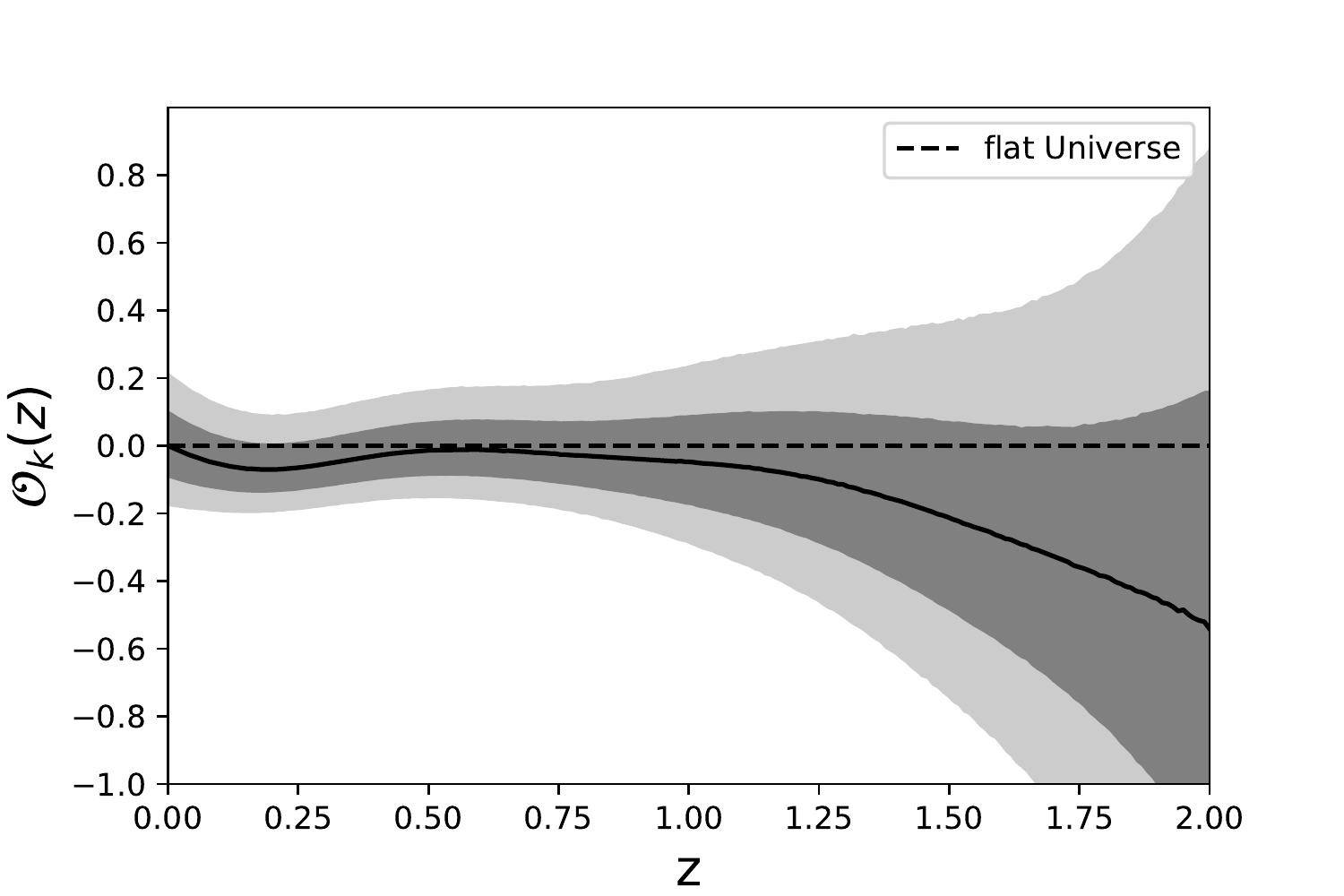}
\caption{The reconstructed $\mathcal{O}_k(z)$ from the SNe Ia and CCH data. The shaded areas are the 1$\sigma$ and 2$\sigma$ confidence regions.}
\label{Ok1}
\end{figure}

With the reconstructed smooth functions for $H(z)$ and $D(z)$,
we derive $\Omega_{k0}$ and $\mathcal{O}_k(z)$
from Eqs. \eqref{ok1} and \eqref{ok2} and the results are shown in Figs. \ref{Ok} and \ref{Ok1}.
The results show that a spatially flat universe is consistent with the reconstructed null tests from CCH and SNe Ia data.
The result is consistent with those obtained
by using different smoothing methods \citep{Shafieloo:2009hi,Li:2014yza,Sapone:2014nna,Cai:2015pia,LHuillier:2016mtc,Marra:2017pst,Li:2019bbg}.
The null test \eqref{ok1} for $\Omega_{k0}$
also supports the FLRW metric.
Since the reconstructions of both $E(z)$ and $D(z)$ depend on the value of $H_0$,
so the null tests \eqref{ok1} and \eqref{ok2} are not totally model independent.

\section{Model independent measurement on the cosmic curvature}
\label{sec3}

In this section, we use the reconstructed Hubble parameter and the
observational data of SNe Ia and redshift space distortions to constrain
the cosmic curvature.
With the smooth function $H(z)$ reconstructed from the CCH data,
we use a simple trapezoidal rule \citep{Holanda:2012ia}
to calculate the proper distance,
\begin{equation}
\label{integral}
\begin{split}
 d_p(z)=&\int_0^z\frac{dz'}{H(z')}\\
 \simeq&\frac{1}{2}\sum_{i=0}^n(z_{i+1}-z_i)\left[\frac{1}{H(z_{i+1})}+\frac{1}{H(z_i)}\right].
\end{split}
\end{equation}
We checked the accuracy of the above approximation with $\Lambda$CDM model and found good agreement.
Using the standard error propagation formula, we obtain
the error of the proper distance
\begin{equation}
\sigma_{dp}^2=\sum_{i=0}^ns_i^2,
\end{equation}
where
\begin{equation}
s_i=\frac{1}{2}(z_{i+1}-z_i)\left(\frac{\sigma_{H_{i+1}}^2}{H_{i+1}^4}
+\frac{\sigma_{H_{i}}^2}{H_{i}^4}\right)^{1/2}.
\end{equation}
Through the GP method and the above integration,
we obtain the smooth function of the proper distance $d_p(z)$ and its error $\sigma_{dp}(z)$ from the CCH data. Note that no specific cosmological
model is assumed, so the reconstructed $d_p(z)$ is model independent.

\subsection{The measurement form SNe Ia data}
Using the reconstructed result for $d_p(z)$, we calculate
the luminosity distance
\begin{equation}
d_L(z)=\begin{cases}
               \frac{c(1+z)}{H_0\sqrt{-\Omega_{k0}}}\sin[H_0\sqrt{-\Omega_{k0}} d_p(z)], & \Omega_{k0}<0, \\
               c(1+z)d_p(z), &\Omega_{k0}=0, \\
               \frac{c(1+z)}{H_0\sqrt{\Omega_{k0}}}\sinh[H_0\sqrt{\Omega_{k0}} d_p(z)], & \Omega_{k0}>0.
              \end{cases}
\end{equation}
The reconstructed luminosity distance now takes the role of a theoretical model.
The parameters $\Omega_{k0}$ and $H_0$ appear in the combination of $\Omega_{k0} H_0^2$
in the luminosity distance, so they are degenerate and we can only constrain
the parameter $\Omega_{k0} h^2$. Even though we don't know the exact value
of $\Omega_{k0}$, the value of $\Omega_{k0} h^2$ is enough to tell us
whether the Universe is spatially flat, open or closed and the conclusion
is independent of the value of the Hubble constant.
The error $\sigma_{dL}$ of the reconstructed luminosity distance is
 \begin{equation}
 \sigma_{dL}(z)=\begin{cases}
 c(1+z)\cos[H_0\sqrt{-\Omega_{k0}} d_p(z)]\sigma_{dp}(z), & \Omega_{k0}<0, \\
 c(1+z) \sigma_{dp}(z), &\Omega_{k0}=0, \\
 c(1+z)\cosh[H_0\sqrt{\Omega_{k0}}d_p(z)]\sigma_{dp}(z), & \Omega_{k0}>0,
 \end{cases}
 \end{equation}
and the error $\sigma_{\mu}$ of the reconstructed distance modulus $\mu_{gp}=5 \log_{10}(d_L/\text{Mpc})+25$ is
\begin{equation}
\sigma_{\mu}=\frac{5}{\ln10}\frac{\sigma_{dL}}{d_L}.
\end{equation}
The error is added to the observational error in quadrature
and the total error of the distance modulus is $\Sigma_{\mu}=\Sigma_{obs}+\sigma_{\mu}^2$.
Now we fit the only parameter $\Omega_{k0} h^2$ to
the 1048 Pantheon sample of SNe Ia by minimizing
the $\chi^2$ function
\begin{equation}\label{chisn}
\chi_{SN}^2=\Delta\mu^T\cdot \Sigma_{\mu}^{-1}\cdot\Delta\mu,
\end{equation}
where $\Delta\mu=\mu_{obs}-\mu_{gp}$,
and the result is $\Omega_{k0}h^2=0.102 \pm 0.066$.
The same
approach was used in \cite{Wei:2016xti}, but they
reconstructed $E(z)$ and $H_0 d_p$, so the value of the Hubble
constant plays a significant role in the determination of the cosmic curvature.
Due to the uncertainty or the model dependence of the value of $H_0$
and the arbitrary normalization
in the SNe Ia data, as well as the degeneracy between $\Omega_{k0}$
and $H_0$, the approach in \cite{Wei:2016xti} has drawbacks like
model dependence.
However, the model independent reconstruction of $H(z)$
and $d_p(z)$ presented above and the constraint on $\Omega_{k0} h^2$
obtained here do not suffer the $H_0$ problem.

\subsection{The measurement from redshift space distortions}

The growth of large structure can not only probe the background evolution
of the Universe, but also distinguish modified theories of gravity.
In this subsection, we propose a model independent method
to use the growth rate data measured from
RSD to constrain the spatial curvature.

\begin{figure}
\includegraphics[width=0.9\linewidth]{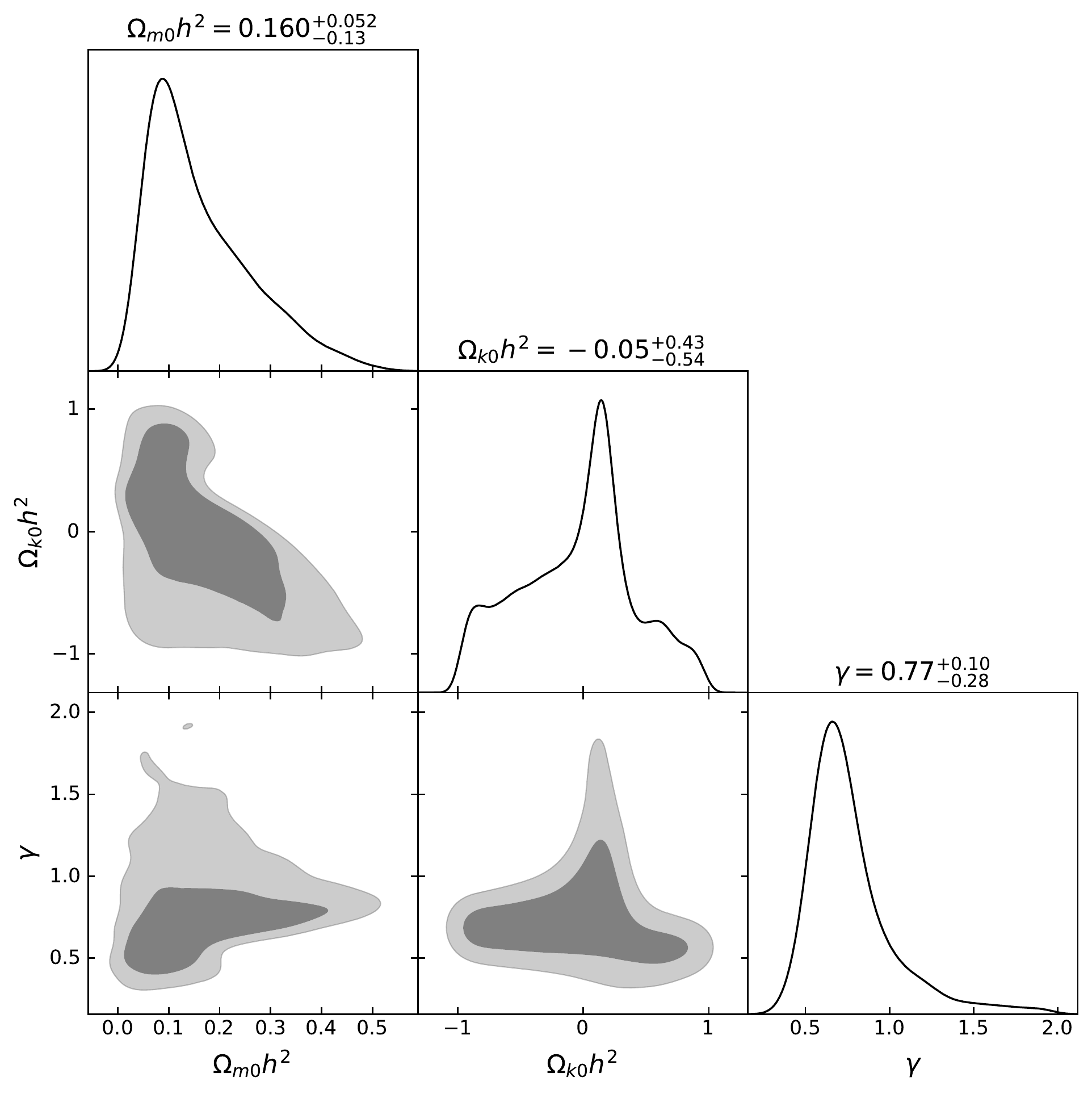}
\caption{The  marginalized  likelihood distributions for fitting the RSD and CCH data.}
\label{HRSD}
\end{figure}

To the linear order of perturbation, the matter density perturbation $\delta=\delta\rho_m/\rho_m$ satisfies the following equation
\begin{equation}
\label{eq1}
\ddot\delta+2H\dot\delta-4\pi G_{eff}\rho_m\delta=0,
\end{equation}
where $\rho_m$ is the background matter density and $G_{eff}$ denotes the effect of modified gravity. For Einstein's general relativity, $G_{eff}$ is Newton's gravitational constant $G$.
Using the growth factor $f(a)=d\ln\delta/d\ln a$, a good
approximated solution to Eq. \eqref{eq1} is \citep{Gong:2009sp}
 \begin{equation}
 f(z)=\Omega_m(z)^\gamma+(\gamma-4/7)\Omega_k(z),
 \label{5}
 \end{equation}
where
\begin{equation}
\label{7}
\Omega_m(z)=\frac{\Omega_{m0}(1+z)^3}{(H/H_0)^2}=\frac{10^4\Omega_{m0} h^2 (1+z)^3}{H^2(z)},
\end{equation}
\begin{equation}
\label{8}
\Omega_k(z)=\frac{\Omega_{k0}(1+z)^2}{(H/H_0)^2}=\frac{10^4\Omega_{k0} h^2 (1+z)^2}{H^2(z)},
\end{equation}
the subscript 0 represents the current value of the variables and
the growth index $\gamma$ depends on the model and gravitational theory.
Note that there are three parameters $\Omega_{m0} h^2$, $\Omega_{k0} h^2$
and $\gamma$ in Eq. \eqref{5}.
Since the matter density $\Omega_m(z)$ and the cosmic curvature $\Omega_k(z)$
can be obtained from the smooth function $H(z)$ reconstructed from the CCH data,
so we can combine the parametrization \eqref{5}
and the growth rate data to constrain the present cosmic curvature.
Note that this method constrains not only the cosmic curvature but
also the growth index $\gamma$ which is an indicator of underlying theory or model.
We use 35 RSD data points compiled in \cite{zhang:2018gjb}
to constrain the cosmic curvature. The RSD data measure $f\sigma_8(z)$,
\begin{equation}
 f\sigma_8(z)=f(z)\sigma_{8,0}\delta(z)/\delta_0,
\label{3}
\end{equation}
where $\sigma_8(z)$ is the matter power spectrum normalization on the scale of $8h^{-1}$Mpc.
Substituting Eqs. \eqref{5}, \eqref{7}, \eqref{8} into Eq. \eqref{3}, we get
\begin{equation}
\label{9}
\begin{split}
f\sigma_8(z)=&\sigma_{8,0}[\Omega_m^\gamma+(\gamma-4/7)\Omega_k]\\
&\times
\exp\left(-\int_0^z\frac{[\Omega_m^\gamma+(\gamma-4/7)\Omega_k]}{1+z'}dz'\right).
\end{split}
\end{equation}
Following the same error propagation procedure discussed in the previous section, we
estimate the error $\sigma_{fg}$ on the reconstructed $f\sigma_8$ from the error in $H(z)$ and added it to the observational error in quadrature.
The three parameters $\Omega_{k0}h^2$, $\Omega_{m0} h^2$ and $\gamma$ are
then fitted to the 35 RSD data points with the $\chi^2$ minimization.
The results are shown in Fig. \ref{HRSD} and Table \ref{table1}.
Due to the large uncertainty, both the flat $\Lambda$CDM and DGP models are consistent
with the combined RSD and CCH data.

\begin{figure}
\includegraphics[width=0.9\linewidth]{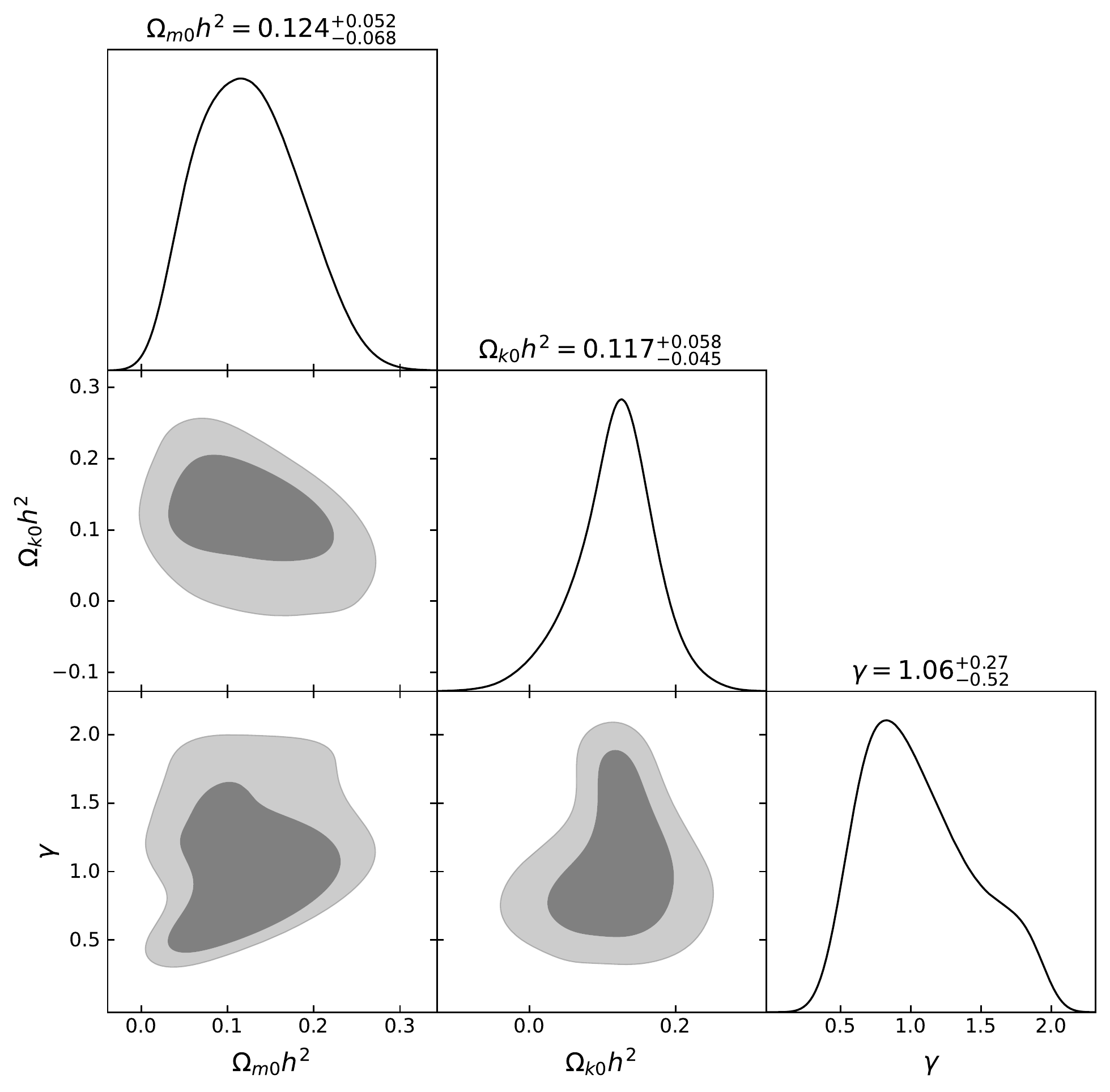}
\caption{The  marginalized  likelihood  distributions
for fitting the combined RSD, SNe Ia and CCH data.}
\label{HRSDDL}
\end{figure}

Finally, we fit the three parameters $\Omega_{k0}h^2$,
$\Omega_{m0} h^2$ and $\gamma$ to the combined
SNe Ia, RSD and CCH data
and we obtain $\Omega_{m0} h^2=0.124^{+0.052}_{-0.068}$,
$\Omega_{k0}h^2=0.117^{+0.058}_{-0.045}$ and $\gamma=1.06^{+0.27}_{-0.52}$.
The results are shown in Fig. \ref{HRSDDL} and Table \ref{table1}.
With the addition of SNe Ia data, the constraint on the cosmic curvature
becomes tighter and the combined data favor a spatially open universe
at almost $2\sigma$ confidence level. However,
no evidence of deviation from general relativity is detected
although DGP model is also consistent with the combined data.

\begin{table}
\centering
\begin{tabular}{cccc}
\hline\hline
data & $\Omega_{k0}h^2$ & $\Omega_{m0}h^2$ & $\gamma$   \\
\hline
 SN+H(z) & $0.102\pm0.066$ & &  \\
 RSD+H(z) & $-0.05_{-0.54}^{+0.43}$ & $0.160_{-0.13}^{+0.052}$ & $0.77_{-0.28}^{+0.10}$  \\
 SN+RSD+H(z) & $0.117_{-0.045}^{+0.058}$ & $0.124_{-0.068}^{+0.052}$ & $1.06_{-0.52}^{+0.27}$ \\
 \hline\hline
\end{tabular}
\caption{The results for the constraint.}
\label{table1}
\end{table}

\section{Conclusions}
\label{sec4}

Based on the relation between distances and the Hubble expansion rate
derived from the background FLRW metric, the null test \eqref{ok1} of $\Omega_{k0}$
not only determines the cosmic curvature of the Universe but also
provides a consistency test for the FLRW metric, while the null test \eqref{ok2} probes
the flatness of the Universe only.
The null tests \eqref{ok1} and \eqref{ok2} are model independent
because neither a particular cosmological model nor
a gravitational theory is used.
Using the GP method, the luminosity distance $d_L(z)$ and the dimensionless comoving distance $D(z)$
are reconstructed from SNe Ia data,
and the Hubble expansion rate $H(z)$ is reconstructed from the CCH data.
Note that the reconstruction of $D(z)$ from SNe Ia data depends  on the value of Hubble constant.
The null tests \eqref{ok1} and \eqref{ok2}
with the reconstructed $H(z)$ and $D(z)$
find no deviation from a spatially flat universe
and FLRW metric is consistent with the observational data.
The results also imply that there is
no tension between SNe Ia and CCH data.

The null tests determine the values of $\Omega_{k0}$
and $\mathcal{O}_k$ at each redshift
and check whether it is a constant or zero.
Alternatively, without reconstructing the luminosity distance
we can determine $\Omega_{k0}$ from SNe Ia data by
using the $\chi^2$ minimization. This way
of determining the cosmic curvature relies on the FLRW
metric only and therefore is model independent.
Combining the SNe Ia
and CCH data, we find that $\Omega_{k0}h^2=0.102 \pm 0.066$
and this result is consistent with a spatially flat universe
at the $2\sigma$ confidence level.
Since we don't use the value of the Hubble constant,
this conclusion avoids the problem of the Hubble constant.
In addition to the distance data which depends
on the background geometry, we also propose a novel
model independent method to use the RSD data which measure
the growth of large structure to determine the cosmic curvature.
This method constrains not only the cosmic curvature but
also the growth index $\gamma$, so we can also distinguish
the underlying theory.
The constraint from the combined CCH, SNe Ia and RSD data is
$\Omega_{k0}h^2=0.117^{+0.058}_{-0.045}$ and $\gamma=1.06^{+0.27}_{-0.52}$.
While no evidence of cosmic curvature is found from the combined
CCH and RSD data, we find a deviation from a spatially flat universe
at almost $2\sigma$ confidence level from the combined CCH,
SNe Ia and RSD data.
The results suggest that SNe Ia data prefers an open universe.
No  deviation from $\Lambda$CDM model is found from
the the combined CCH, SNe Ia and RSD data
and the DGP model is also consistent with the combined observational data.
More accurate data in the future may help
resolve the issue of the spatial curvature of the Universe.
	
\section*{acknowledgements}
This research was supported in part by the National Natural Science
Foundation of China under grant no. 11875136 and
the Major Program of the National Natural Science Foundation of China under grant no. 11690021.


\section*{Data availability}
The data underlying this article are available in the paper and the cited references.

\end{document}